\documentclass[a4paper]{article}

\usepackage{INTERSPEECH2022}
\usepackage{amsmath, hyperref}
\usepackage{cite}
\usepackage{xcolor}

\def\s{{\mathbf s}}

\def\L{{\cal L}}
\def\E{\mathbf E}
\def\S{\mathbf S}
\DeclareMathOperator*{\argmin}{arg\,min}
\def \scaledERM         {\overline{\IRM}}

\def \IRM               {\mathbf M}           
\def \ERM               {\widehat{\IRM}}   

\def\Mhat{\widehat{\mathbf M}}

\title{A Conformer-based Waveform-domain Neural Acoustic Echo Canceller Optimized for ASR Accuracy}
\name{Sankaran Panchapagesan, Arun Narayanan, Turaj Zakizadeh Shabestary, Shuai Shao, Nathan Howard, Alex Park, James Walker, Alexander Gruenstein}
\address{Google LLC, U.S.A.}
\email{\{panchi,arunnt\}@google.com}

\begin{document}

\maketitle
\begin{abstract}
Acoustic Echo Cancellation (AEC) is essential for accurate recognition of queries spoken to a smart speaker that is playing out audio. Previous work has shown that a neural AEC model operating on log-mel spectral features (denoted ``logmel" hereafter) can greatly improve Automatic Speech Recognition (ASR) accuracy when optimized with an auxiliary loss utilizing a pre-trained ASR model encoder. In this paper, we develop a conformer-based waveform-domain neural AEC model inspired by the ``TasNet” architecture. The model is trained by jointly optimizing Negative Scale-Invariant SNR (SISNR) and ASR losses on a large speech dataset. On a realistic re-recorded test set, we find that cascading a linear adaptive AEC and a waveform-domain neural AEC is very effective, giving 56-59\% word error rate (WER) reduction over the linear AEC alone. On this test set, the 1.6M parameter waveform-domain neural AEC also improves over a larger 6.5M parameter logmel-domain neural AEC model by 20-29\% in easy to moderate conditions.
By operating on smaller frames, the waveform neural model is able to perform better at smaller sizes and is better suited for applications where memory is limited.

\end{abstract}
\noindent\textbf{Index Terms}: Acoustic echo cancellation, TasNet, Waveform Neural AEC, sequence-to-sequence model, multi-task loss

\section{Introduction}
\label{sec:intro}

We consider the problem of recognizing queries spoken to a smart speaker 
device that is playing out audio such as text-to-speech (TTS) responses, audiobooks or music. In such conditions, the received device microphone signal $x(n)$ contains a mixture of the target speech query $s(n)$, an echoed version $e(n)$, of the device playback reference audio $r(n)$, and background noise $z(n)$:
\begin{equation}
x(n) = s(n) + e(n) + z(n)
\end{equation}
In conventional AEC, a finite impulse response linear adaptive filter $\hat{h}(n)$ is usually used to approximate the relationship between $r(n)$ and $e(n)$, which includes the combined loudspeaker and room echo responses to device playback \cite{HanslerSchmidt2005}. $\hat{h}(n)$ is estimated from some time period before the speech query starts, when $s(n)=0$, and then used to estimate and cancel the echo.
The enhanced speech query is then:
\begin{eqnarray}
\hat{s}(n) &=&  x(n) - \hat{h}(n)*r(n) \nonumber \\
 &=& s(n) + [e(n)-\hat{h}(n)*r(n)] + z(n)
\end{eqnarray}
The assumption that the room echo response is linear, is usually a reasonably good one, and linear AEC methods are quite effective at improving the signal-to-interference ratio (SIR) of the speech input to a smart speaker that is playing out audio.

There is still residual echo at the output of the linear finite length AEC due to long room reverberation times, or non-linearities in the device loudspeaker and the room response.
Residual echo and noise can cause significant degradation in speech quality and intelligibility, and hinder speech or hotword recognition.
Several neural network based methods have recently been proposed for AEC and residual echo suppression, including architectures based on LSTMs, dilated convolutions, attention networks, U-Nets and others \cite{Zhang2018, Lei2019, Fazel2019, Fazel2020, chen2020nonlinear, kim2020attention, ma2021echofilter, westhausen2021acoustic, ma2021multi, chen2021nonlinear, watcharasupat2021end}. 
However, almost all of these approaches are focused on optimizing and improving enhancement and speech quality metrics, which are not matched with our goal of speech recognition.

Recently there have been some works focused on AEC for speech or hotword recognition \cite{howard2021neural, cornell2021implicit}. In \cite{howard2021neural}, a neural AEC model operating on logmel  features was proposed, together with a novel {\em ASR loss} that improved ASR accuracy. The AEC model was an LSTM-based sequence-to-sequence model taking mixture and reference features as input and predicting enhanced target speech features. In addition to a spectral L1+L2 loss between predicted and target logmel features, a pre-trained  ASR model was used to compute the ASR loss, defined as the L2 loss between the two sets of encoder output representations obtained with target and predicted features as inputs.  In an ablation experiment, removing the ASR loss degraded WER by 12-17\%. In \cite{cornell2021implicit}, an implicit AEC approach was used, taking mixture and reference features as input to a temporal convolution network for predicting posterior probabilities related to the target tasks of either multi-keyword-spotting or device-directed speech detection. Although narrowly focused on the tasks of interest, this implicit AEC approach was shown to be more parameter efficient than first predicting enhanced features.

In this paper, we develop a conformer-based \cite{gulati2020conformer} waveform-domain neural AEC model inspired by the TasNet architecture \cite{luo2018tasnet}, and optimize it for ASR accuracy using the ASR loss from \cite{howard2021neural}. By operating on smaller 5ms signal windows, we are able to obtain better performance than a larger 6.5M parameter logmel-domain neural AEC model, using a smaller 1.6M parameter waveform-domain neural AEC model. This model is therefore better suited for applications where memory is limited, e.g. on mobile devices. A waveform-domain model also has the advantage that it can be simultaneously optimized for improved  ASR accuracy, hotword accuracy, and speech quality and intelligibility for human listening, using multiple loss functions at different time scales.
Thus the model can be targeted for multiple applications without needing to retrain backend models, unlike implicit AEC approaches. With batched computation using specialized machine learning accelerator chips, any additional computation due to a higher frame rate might be handled without additional latency.

The rest of the paper is organized as follows. Section \ref{sec:related-work} discusses related work. In Section \ref{sec:wave-neural-aec}, the architecture of the proposed waveform-domain neural AEC model is described, together with the training loss. In Section \ref{sec:expts}, the experimental setup including training and evaluation data are described. Experimental results are presented in Section \ref{sec:results}, and conclusions and ideas for future work in Section \ref{sec:conclusions}.

\vspace{-0.05in}
\section{Related Work}
\label{sec:related-work}

In \cite{chen2020nonlinear}, a Conv-TasNet model was proposed for residual echo suppression, and trained on input reference and residual signals from a linear AEC. In \cite{ma2021echofilter}, EchoFilter, a TasNet-style masking model using attention and LSTM modules was proposed for AEC, along with an auxiliary double-talk detection network which was found to be useful. In CAD-AEC \cite{Fazel2020}, a cascade of a linear adaptive AEC and a neural model was trained, with the neural model taking log-spectra of mixture, reference and output error signal of the linear AEC as inputs. The neural model consisted of a recurrent encoder layer, a contextual attention module containing two residual MHSA layers, and a recurrent decoder layer. As mentioned in Section \ref{sec:intro}, the above neural AEC models were optimized only for enhancement and speech quality, unlike our model which also targets speech recognition accuracy.
In \cite{koizumi2021df}, the conformer architecture was extended with stacked 1-D dilated depthwise convolutions, and used with TasNet for speech enhancement. In \cite{OMalleyJointConformerFrontEnd2021}, a joint conformer-based frontend was developed to handle additional contextual input signals for AEC, multi-channel enhancement and ID-based speaker separation.  The model operated on logmel features, and predicted enhanced logmel features.

This paper, to the best of our knowledge, is the first to apply a conformer-based TasNet architecture to AEC, and is the first waveform-domain AEC model optimized for ASR accuracy. 

\vspace{-0.05in}
\section{Waveform-domain Neural AEC Model}
\label{sec:wave-neural-aec}

\subsection{Model Architecture}
\vspace{-0.05in}

\begin{figure}[t]
  \centering
\includegraphics[width=0.95\linewidth]{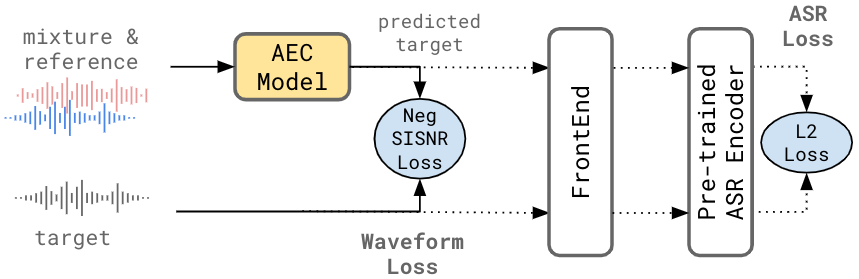}
\vspace{-0.05in}
  \caption{High-level block diagram of waveform-domain neural AEC model, with Negative SISNR and ASR training losses.}
  \label{fig:Losses-Waveform-Neural-AEC}
\vspace{-0.15in}
\end{figure}

Figure \ref{fig:Losses-Waveform-Neural-AEC} shows the high-level block diagram of the waveform-domain neural AEC system. The AEC model takes the mixture (i.e., the microphone signal) and reference waveforms, $x(n)$ and $r(n)$ respectively, as input and produces an enhanced waveform $\hat{s}(n)$ as output. The model is trained with a combination of Negative SISNR loss and the ASR loss proposed in \cite{howard2021neural}. The FrontEnd that computes logmel features needed for the ASR loss, is implemented as a non-trainable layer that backpropagates gradients from the ASR loss layer during model training.

\begin{figure}[t]
  \centering
\includegraphics[width=0.95\linewidth]{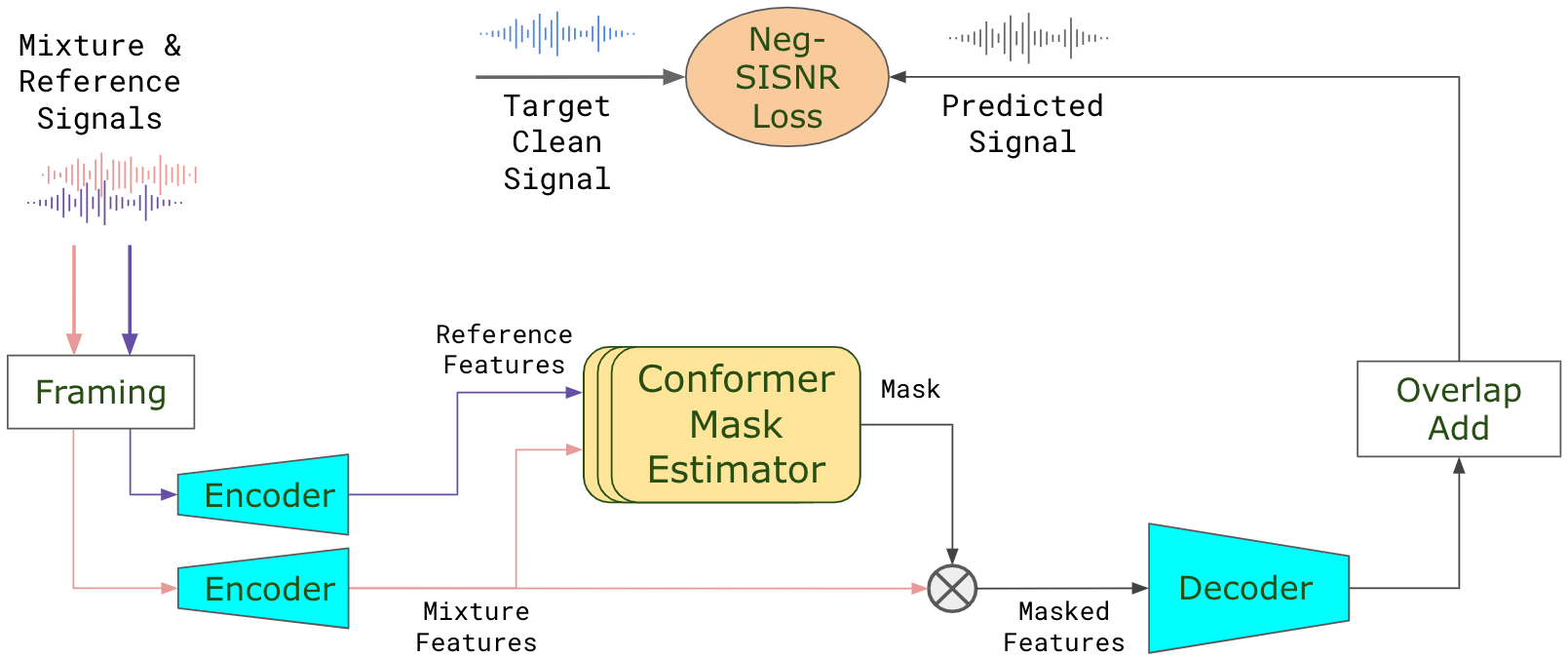}
\vspace{-0.05in}
  \caption{Waveform-domain neural AEC architecture with ASR loss omitted for clarity.}
  \label{fig:TasNet-AEC}
\end{figure}

Figure \ref{fig:TasNet-AEC} shows our proposed architecture for the neural AEC model inspired by TasNet \cite{luo2018tasnet}, with the ASR loss  omitted for clarity. The model performs enhancement by masking in a learned feature space. The input mixture and reference signals are framed, and converted to features by separate linear encoder layers. The mixture and reference features are stacked and input into a mask estimator network which estimates an enhancement mask in the feature domain. The mask is multiplied with the mixture features to produce predicted features, which are converted into the predicted waveform by the linear decoder layer followed by the overlap-add operation. 

\begin{figure}[t]
  \centering
\includegraphics[width=0.99\linewidth]{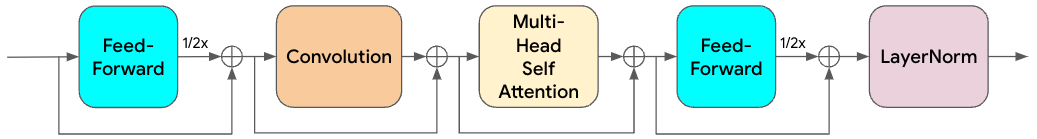}
\vspace{-0.05in}
  \caption{Conformer Layer Block.}
  \label{fig:Conformer-block}
\vspace{-0.15in}
\end{figure}

As noted in Figure \ref{fig:TasNet-AEC}, we use a conformer model \cite{gulati2020conformer} for the enhancement mask estimation. Conformers combine convolutional and transformer models to efficiently model both global and local dependencies in audio signals. The conformer layer block, shown in Figure \ref{fig:Conformer-block}, is the improved one used in \cite{li2021better, narayanan2021cross}, where the order of convolution and multi-head self attention (MHSA) in \cite{gulati2020conformer} have been swapped. This helps avoid the use of relative positional embedding in the attention. In order for the model to be streaming, only previous frame context is used in all convolution, MHSA and normalization modules. As seen in Figure \ref{fig:Conformer-block}, the sub-layer modules all use residual connections, and layer normalization is used after each conformer block. There are two half-step feed-forward networks sandwiching the convolution and MHSA modules. These feed-forward networks internally contain two linear layers separated by a non-linearity. The first linear layer expands the dimension by a factor of 4, and the second projects back down to the model dimension. The convolution module uses point-wise convolutions, gated linear units, 1-D depth-wise convolution, and group normalization. Further details can be found in \cite{gulati2020conformer, li2021better, narayanan2021cross}.

\subsection{Loss Functions}
\vspace{-0.05in}

As mentioned above, the total training loss is:
\begin{eqnarray}
\L &=& -SISNR(\s, \hat{\s}) + \lambda \L_{ASR}
\label{eqn:loss}
\end{eqnarray}
where $\lambda$ is a hyperparameter. 
SISNR is defined as the following SNR obtained after scaling the target signal to have least squares error with the predicted signal \cite{luo2018tasnet, le2019sdr}:
\begin{equation}
    SISNR(\s, \hat{\s}) = 10 \log_{10} \frac{\lVert\alpha^*\s\rVert^2}{\lVert\alpha^*\s-\hat{\s}\rVert^2}
\end{equation}
\vspace{-0.05in}
where
\vspace{-0.05in}
\begin{equation}
\displaystyle \alpha^* = \argmin_\alpha \lVert\alpha\s-\hat{\s}\rVert^2 = \frac{\langle\s,\hat{\s}\rangle}{\lVert\s\rVert^2}
\end{equation}

$\L_{ASR}$ is defined as an L2 loss between target and predicted ASR encoder output sequences:
\begin{equation}
 \L_{ASR} = \sum_{k} \left\lVert \E_{ASR}(\S_k) - \E_{ASR}(\hat{\S}_k) \right\rVert_2^2
\vspace{-0.05in}
\end{equation}
where $\S_k$ and $\hat{\S}_k$ are feature vectors at frame $k$ computed by the frontend from the target and predicted signals respectively, and $\E_{ASR}(\cdot)$ is the ASR encoder function. For the experiments in this paper, we used the conformer encoder from the ASR model described in  \cite{li2021better}.

\vspace{-0.05in}
\section{Experiments}
\label{sec:expts}

\subsection{Training Data}
\vspace{-0.05in}
The target speech utterances for training the neural AEC models consists of utterances from Librispeech \cite{panayotov2015librispeech} (960 hours),  LibriVox\footnote{\url{https://librivox.org}} (46.5k hours), and internal vendor collected datasets containing utterances both with and without the hotword (3.3k hours), giving a total of 50.8k hours. Following \cite{howard2021neural}, we use both synthetic and real echoes to create  mixture signals. Real echoes are obtained by rerecording utterances from Librispeech and from an internal dataset collected for training Text-to-Speech (TTS) models on a Google Home device from multiple rooms. The latter was chosen since an important use case is to cancel TTS responses played by the device. For synthetic echoes, we use utterances from Librispeech and Librivox, and noise snippets from Getty\footnote{\url{https://www.gettyimages.com/about-music}} and YouTube Audio Library\footnote{\url{https://youtube.com/audiolibrary}}. These are convolved with synthetic room impulse responses (RIRs), with reference location close to the microphones to mimic device playback. Target speech is also convolved with RIRs to simulate farfield conditions, and mixed with real or synthetic echoes at signal-to-echo ratio (SER) between -20 dB and 5 dB.

\subsection{Simulated Librispeech Test Set}
\label{sec:librispech-test-set}
\vspace{-0.05in}

Simulated Librispeech test sets are created from the Librispeech test-clean subset by artificially adding reverberation and noise to the target speech queries, and adding held out re-recorded echoes at SERs of 5 dB, 0 dB, -5 dB and -10 dB. Similar to the training utterances, each test utterance also had the associated original playback reference signal during evaluation. The test sets contained 2620 utterances with around 52.5k words each. 

\subsection{Re-recorded Test Set}
\label{sec:rerecorded-test-set}
\vspace{-0.05in}

We also created a more realistic test set by combining re-recorded speech queries and re-recorded reference signals. The speech queries were recorded from three different speaker positions at distances of 1.3m, 3.3m and 5.2m from the device. For re-recording the reference, the device playback volume setting was varied over a set of values including a low value of 3 and the range 6-10 (where 10 was the maximum setting). Each test set utterance begins with around 10 sec of echoed reference only, followed by the target speech query mixed with interfering echoed reference, and ends with several seconds of continuing echoed reference, with a total length up to 30 sec. The test set was partitioned based on playback volume and speaker position into Easy, Moderate and Difficult sets, containing respectively around 19k, 12.3k, and 11.4k utterances, and around 141k, 92k, and 85k words.
This test set is a very challenging one for the neural AEC models due to significant mismatch between the test set and the simulated data used to train the models.

\subsection{Linear Adaptive AEC}
\vspace{-0.05in}
We also use a Linear AEC system that performs subband adaptive filtering using STFT similar to \cite{avargel2006}, but uses longer STFT frames (128ms) and within-band only filters of order 4.

\subsection{Logmel-domain Neural AEC}
\vspace{-0.05in}

In this paper, our baseline was an improved version of the logmel neural AEC model \cite{howard2021neural}, proposed in \cite{Narayanan2022MaskScalar}.
Instead of directly predicting enhanced logmel features as in \cite{howard2021neural}, this model predicts enhancement masks in the logmel domain, with ideal ratio masks as targets. The predicted masks $\ERM$ are post-processed
to trade-off residual noise and distortion:
\begin{equation}
\scaledERM = \max(\Mhat^\alpha, \beta), \label{eq:scale}\\
\end{equation}
where $\alpha \in [0, 1]$ exponentially scales the mask, and $\beta$ floors the mask. While tuning a fixed $\alpha$ for best WER works well, further ASR accuracy improvements were obtained by predicting $\alpha$ from encoded input features on a per-frame basis using a separate mask scalar prediction network, and using only ASR loss to train this network. The final enhanced logmel features were obtained by multiplying input noisy features with predicted masks. This logmel feature masking approach was shown to perform significantly better than direct enhanced logmel prediction for both multichannel enhancement and AEC tasks.

The trained masking-based logmel-domain neural AEC model had an encoder-decoder architecture, with a conformer model for the encoder and a linear layer for the decoder. The conformer model had 4 layers with model dimension 256, convolutional blocks with kernel size of 15, and causal attention with a left context of 31 frames. 128-dimensional logmel features were computed over 32ms frames with 10ms shift. The model input sequence consisted of 512-dimensional vectors obtained by stacking 4 frames of logmel features, and then subsampling by a factor of 3. The linear decoder layer converted 256-dim. encoded features to 512-dim. output enhancement masks. Masked 512-dim. features are used as input to the ASR model.
The total size of the model was 6.5M parameters.

\subsection{Waveform-domain Neural AEC}
\vspace{-0.05in}
The input signals to the waveform-domain neural AEC model are framed using windows of length 5ms (80 samples at 16kHz sampling rate), shifted by 2.5ms (40 samples). 
The learned feature dimension was 128, which was also the dimension of the 4 conformer layers of the mask estimator. The convolutional blocks in the conformer layers had a kernel size of 15, while the causal attention had 8 heads with a left context of 31 frames.
The linear decoder layer used tanh activation to produce audio samples in the range $(-1, 1)$.
The total size of the model was 1.6M parameters.

Taking the 31 frames left context of the attention in the four conformer layers, and the frame shift of 2.5ms, we compute that the waveform-domain neural AEC model uses a total past context of approximately $4\times31\times2.5 = 310$ms. The logmel-domain neural AEC model uses a significantly longer past context of $4\times31\times30 = 3720$ms.

We used the Lingvo toolkit \cite{Shen2019Lingvo} to train models. During training, the ASR loss weight ($\lambda$ in Equation \ref{eqn:loss}) is increased linearly, starting from $0.0$ at 5k steps to the selected value at 20k steps, and kept fixed after that.

\subsection{ASR Evaluations}
\vspace{-0.05in}
The ASR model used for evaluations is a recurrent neural net transducer with an LSTM-based encoder \cite{sainath2020streaming}. Similar to the logmel AEC models, it is trained on 128-dimensional logmel features computed for 32 msec windows with 10 msec hop. Features from 4 contiguous frames are stacked to create a 512-dimension representation, that is subsampled by a factor of 3. The model is trained using $\sim$400k hours of English speech from domains like VoiceSearch, YouTube, Telephony and Farfield. Utterances are anonymized and hand-transcribed for training. The model also uses data augmentations like SpecAug \cite{park2019specaugment} and simulated noise \cite{kim2017mtr}. Note that the ASR model is not jointly trained with the AEC model, and is kept frozen during training and inference. For inference, we use label synchronous beam search, with no endpointer.

\vspace{-0.05in}
\section{Results}
\label{sec:results}
\subsection{Results on simulated Librispeech test sets}

We first evaluated the different AEC methods on the simulated Librispeech test set described in Section \ref{sec:librispech-test-set}. 
This well-matched test was used to measure the effectiveness of the ASR loss, to tune the ASR loss weight ($\lambda$ in Equation \ref{eqn:loss}), and to compare the logmel and waveform neural AEC models.

Table \ref{table:vary-asr-loss-weight} shows the effect of varying $\lambda$ on the performance of the waveform neural AEC model. We evaluate both the enhancement performance in terms of SISNR improvement (SISNRi), and the speech recognition WER.
SISNRi was measured on a development subset containing 1270 utterances from the Librispeech test set, with simulated SER of 0 dB. It can be seen from Table \ref{table:vary-asr-loss-weight} that as $\lambda$ is increased there is a trade-off between WER and enhancement performance in terms of SISNRi. Comparing $\lambda = $~5e3 and $\lambda = $~5e4, there is a relatively steeper drop of 2.6 dB in SISNRi, with relatively small WER gains at -5 dB, 0 dB and 5dB test SERs, but a larger WER gain at -10 dB test SERs.
We fix the ASR loss weight at $\lambda = $~5e4 in the following experiments.
Comparing the first and last rows, it is clear that the ASR loss gives  significant WER reductions of 7-19\% at 5dB to -10dB SER.

\begin{table}[ht]
\caption{Effect of varying ASR loss weight $\lambda$ on \textnormal{(i)} SISNRi (dB) on 0 dB SER Librispeech test subset and \textnormal{(ii)}  WERs(\%) on different SER Librispeech test subsets.}
\vspace{-0.05in}
\centering
\resizebox{0.9\columnwidth}{!}{%
\begin{tabular}{lccccc}
\toprule
\textbf{$\lambda$}   & \textbf{SISNRi}                               &  \multicolumn{4}{c}{\bfseries WERs (\%)}  \\
   &          (dB)                             & ~5 dB & 0 dB  & -5 dB & -10 dB~       \\
\midrule
0       & 20.9  & 11.0 & 14.0 & 19.4 & 28.8             \\
5e1      & 21.0  & 10.8 & 14.0 & 18.9 & 27.3             \\
5e2     & 20.5  & 10.2 & 12.9 & 17.6 & 26.7            \\
5e3    & 20.1  & ~9.7 & 11.8 & 15.4 & 23.2              \\
5e4   & 17.5  & ~9.8 & 11.5 & 15.3 & 21.9              \\
\bottomrule
\end{tabular}
}
\label{table:vary-asr-loss-weight}
\vspace{-0.05in}
\end{table}

The results of the different AEC methods on the simulated Librispeech test sets are presented in Table \ref{table:wers-librispeech}. It is seen that both the logmel- and waveform-domain neural AEC models perform significantly better than the Linear AEC at all SERs. It should be noted that these test sets are much better matched to the neural AEC models than the Linear AEC, since the utterances are not designed to specifically begin with a reference-only segment where the adaptive filter can converge before the target speech starts. It is also seen that the waveform-domain neural AEC model yields better ASR performance than the logmel-domain neural AEC model, with 6\% lower WERs at 0dB and -5dB SERs, and 12\% lower WER at -10dB SER.

\begin{table}[ht]
\caption{WERs(\%) with different AEC models on simulated Librispeech test subsets at varying SERs.}
\centering
\resizebox{\columnwidth}{!}{%
\begin{tabular}{lcccc}
\toprule
AEC Method               & 5 dB & 0 dB  & -5 dB & -10 dB  \\ \midrule
No AEC & $36.1$ & $58.0$ & $72.7$ & $80.5$ \\
Linear Adaptive & 19.1 & 24.9 & 26.8 & 28.8 \\
Logmel Neural & ~9.9 & 12.2 & 16.3 & 25.0 \\
Waveform Neural & ~\textbf{9.8} & \textbf{11.5} & \textbf{15.3} & \textbf{21.9} \\
\bottomrule
\end{tabular}
}
\label{table:wers-librispeech}
\vspace{-0.1in}
\end{table}

\subsection{Results on Re-recorded Test Set with Cascaded System}

\begin{figure}[ht]
  \centering
\includegraphics[width=0.95\linewidth]{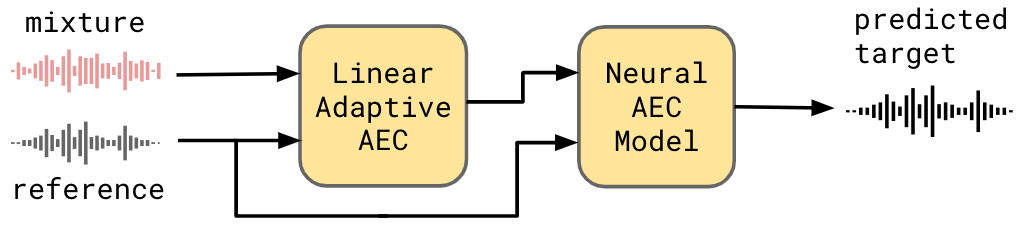}
\vspace{-0.05in}
  \caption{Cascade of Linear Adaptive AEC and Neural AEC.}
  \label{fig:Cascaded-Linear-Neural-AEC}
\vspace{-0.05in}
\end{figure}

We next evaluate the effectiveness of the proposed waveform-domain neural AEC model on the more realistic re-recorded test set described in Section \ref{sec:rerecorded-test-set}. As mentioned there, this test set is a very challenging one for the neural AEC models due to significant mismatch with the simulated data used to train the neural models. Hence, the neural AEC model by itself does not perform well on this test set. However, cascading the Linear Adaptive AEC and the neural AEC model was found to be very effective in improving the WER. This cascaded approach is illustrated in Figure \ref{fig:Cascaded-Linear-Neural-AEC}.

Results of the cascaded approach on the re-recorded test set are presented in Table \ref{table:wers-rerecorded-cascade}. The generally challenging nature of the test set is clear from the poor performance of the Linear AEC alone. Both logmel- and waveform-domain neural AEC models, when cascaded after the Linear AEC, give large improvements in WER over the Linear AEC alone, with the cascade of Linear + Waveform Neural AEC giving WER reductions of 58\%, 59\% and 56\% respectively, on the Easy, Moderate and Difficult test set partitions. It is also seen that the cascade of Linear + Waveform Neural AEC performs better than the cascade of Linear + Logmel Neural AEC, with significant WER reductions of 29\% and 20\% on the Easy and Moderate test set partitions respectively. The Linear + Waveform Neural AEC and the Linear + Logmel Neural AEC systems are comparable on the Difficult test set partition.

\begin{table}[ht]
\caption{WERs on re-recorded test sets with Linear adaptive AEC only, vs cascade of Linear and Neural AEC models, both logmel- and waveform-domain.}
\vspace{-0.05in}
\centering
\resizebox{\columnwidth}{!}{
\begin{tabular}{llccc}
\toprule
AEC Method                  & Easy  & Moderate & Difficult  \\ \midrule
Linear only           & 33.5 & 63.1 & 135.2 \\
Linear + Logmel Neural   & 19.7 & 31.9 & 59.2 \\
Linear + Waveform Neural & \textbf{13.9} & \textbf{25.6} & \textbf{59.1} \\
\bottomrule
\end{tabular}
}
\label{table:wers-rerecorded-cascade}
\vspace{-0.1in}
\end{table}

\vspace{-0.05in}
\section{Conclusions}
\label{sec:conclusions}

In this paper, a waveform-domain neural AEC model was developed with an architecture inspired by the TasNet model, and using conformer layers for the enhancement mask estimation. The model was trained by jointly optimizing Negative SISNR and ASR losses on a large speech dataset simulated with both synthetic and real echoes as interference. Significant reductions in WER of 56-59\% were demonstrated on a realistic re-recorded test set by cascading the waveform-domain neural AEC model after a linear AEC system. Future work would include joint training of the cascade of linear and neural AEC models, efficient model compression to a few hundred thousand parameters, and optimizing the AEC model also for hotword accuracy.

\vspace{-0.05in}
\section{Acknowledgements}
We wish to thank Yuma Koizumi, Tom O'Malley and Joe Caroselli for discussions.

\bibliographystyle{IEEEtran}

\bibliography{mybib}

\end{document}